\documentclass[
 reprint,amsmath,amssymb,aps
]{revtex4-2}

\usepackage{graphicx}
\usepackage{dcolumn}
\usepackage{bm}

\begin{document}

\preprint{APS/123-QED}

\title{Lefschetz-Thimble Method to the Sign Problem \\in the Sachdev-Ye-Kitaev Model}

\author{Yin-Hao Fang}
\altaffiliation[Also at ]{
 Department of Physics, Fudan University, Shanghai 200433, China
}
\author{Yang Qi}
\email{qiyang@fudan.edu.cn}
\affiliation{
 Department of Physics, Fudan University, Shanghai 200433, China
}

\date{\today}

\begin{abstract}
The sign problem in path integrals severely limits the application of quantum Monte Carlo methods to strongly correlated systems, whereas the Lefschetz-thimble method offers a rigorous framework to mitigate this issue. In this paper, we establish a numerical framework based on the Lefschetz thimbles for the Sachdev-Ye-Kitaev (SYK) model. By introducing continuous auxiliary fields via the Hubbard-Stratonovich transformation, the partition function is formulated as a high-dimensional complex integral. We employ the complex Langevin equation to explore the saddle-point distribution of the effective action in the complex plane. Monte Carlo integration is then performed to compute the partition function and Green functions. Our numerical results exhibit excellent agreement with high-temperature expansions, infrared-limit solutions, and determinantal quantum Monte Carlo (DQMC) simulations, confirming the validity of our framework. Furthermore, we demonstrate that our complex Langevin search successfully captures the relevant dominant saddle points governing the SYK model in the large-$N$ limit. Our work deepens the understanding of the sign problem in the SYK model and provides a viable non-perturbative avenue for addressing sign problems in more complex strongly correlated systems.
\end{abstract}

\maketitle

\section{\label{sec:introduction}Introduction}
Within the path integral framework, the evolution of a quantum system is interpreted as a sum over all possible trajectories between the initial and final states \cite{feynman_path, weinbergQFT, peskin2018}. Physical observables are thus formulated as weighted averages over field configurations with the Boltzmann weight $e^{-S_E}$ governed by the corresponding Euclidean action $S_E$ \cite{kardar2007, altland2010}. However, when the action becomes complex or the weight is no longer positive definite, the classical probabilistic interpretation breaks down, and direct sampling methods become inapplicable \cite{gattringer2009, mc_coupled_bf, discrete_hs}. This challenge constitutes the infamous sign problem.

The sign problem hinders the application of quantum Monte Carlo methods to many crucial physical systems \cite{Majorana_fermi_sign, New_sign_QCD, t_J_tensor, Fermion_S_auto_dif, solve_fermion_sign}. In condensed matter physics, the two-dimensional Hubbard model is widely believed to capture the $d$-wave pairing mechanism in cuprate superconductors. Yet in low-temperature and strong-coupling regions, determinant quantum Monte Carlo (DQMC) encounters an exceedingly severe sign problem, leading to an exponential increase in statistical errors \cite{study_2d_hubbard}. In high-energy physics, a similar challenge arises in lattice quantum chromodynamics at finite density, where non-zero chemical potentials introduce complex weights and cause the failure of standard statistical methods \cite{chemical_lattice, C_point_lattice}. Beyond a mere computational obstacle, recent studies also reveal that the sign problem is related to the stability of quantum phases. Specifically, it has been shown that all currently known sign-problem-free models cannot sustain a stable Fermi liquid ground state unless anti-unitary symmetries are spontaneously broken \cite{fermi_liquid_sign_free}.

To mitigate this problem, the Lefschetz-thimble has emerged as a rigorous theoretical framework for non-perturbative physics by complexifying the domain of integration and decomposing a highly oscillatory integral into several stable path integrals with real-valued integrands \cite{Witten2010}. Early studies on low-dimensional fermion systems with Lefschetz thimble have successfully mitigated the sign problem and offered key methodological foundations for more complicated systems \cite{structure_lf_simple_fermion, lf_one_lattice, lef_of_sign}. Extensive implementations established thimble-based Monte Carlo sampling algorithms utilizing Langevin dynamics and Lie derivatives, which were demonstrated to preserve model symmetries and reproduce the results of perturbation theory \cite{new_approach_to_s, mc_on_lef, sign_and_mc_beyond_lef}. Furthermore, this Lefschetz-thimble framework extends well beyond fermion systems, finding successful applications in resolving complicated diffraction patterns, gravitational lensing, formulating Lorentzian path integrals for quantum cosmology, and evaluating highly oscillatory 4-simplex propagators in spin foam models of quantum gravity \cite{born2013, c_quantum_reflections, Lorentzian_q_cosmplogy, spin_foam_q_gravity, numerical_spin_foam, spinform_lf_mc}.

Despite these notable advances, the Lefschetz-thimble method is inherently hindered by the requirement to identify all contributing saddle points and explicitly compute the corresponding intersection numbers, which is computationally demanding for high-dimensional quantum field theories \cite{hybrid_mc_on_lef}. Furthermore, even when these saddle points are fully located, a  residual sign problem emerges \cite{complex_langevin_lef}. This issue arises from the different imaginary parts of the action $\text{Im}\{S\}$ across distinct thimbles, which severely increases computational costs and restricts further applications to smaller lattice volumes \cite{pham2013, chern_simons}. To circumvent these limitations, the Complex Langevin simulation can bypass the explicit structure of global thimbles by naturally navigating the configuration space via stochastic quantization. However, Complex Langevin might converge to incorrect values with a lack of rigorous mathematical guarantees \cite{Aarts_langevin}.

To investigate whether Complex Langevin can correctly identify all contributing saddle points in a larger system, we apply the Lefschetz-thimble method to the Sachdev-Ye-Kitaev (SYK) model. We develop a numerical framework for searching saddle points via the complex Langevin equation, constructing thimble and anti-thimble by gradient flow equations, and performing Monte Carlo (MC) integration on the deformed paths.
The remainder of this paper is organized as follows.
In Sec.~II, we introduce the path integral framework of our statistically equivalent four-body SYK model and outline its high-temperature (HT) expansions and infrared-limit solutions.
In Sec.~III, we detail our numerical implementation of the Lefschetz-thimble method, verify our results in both limits, and analyze the weight distributions of distinct saddle points.
In Sec.~IV, we summarize our findings and discuss future directions.

\section{\label{sec:model}Sachdev-Ye-Kitaev Model}
In this paper, we investigate a $(0+1)$-dimensional SYK model describing $N$ Majorana fermions with independent Gaussian random four-body interactions \cite{spin_fluid_random_H, kitaev1, kitaev2}. The Hamiltonian is defined as:
\begin{eqnarray}
\hat{H} =
&& -\sum_{1\le a<b<c<d\le N} \mathcal{J}_{a b c d}^{(4)} \chi_a \chi_b \chi_c \chi_d \\
=&& - \left [ 
\sum_{1\le a<b\le N} \mathcal{J}_{a b}^{(2)}\chi_a\chi_b 
\right ]^2 - \hat{H}_0,
\label{eq:Hamiltonian}
\end{eqnarray}
where the trivial offset is defined as $\hat{H}_0 = \sum_{a<b} [ \mathcal{J}_{a b}^{(2)} ]^2$. All cross-terms with overlapping indices (e.g., $a=c$ or $a=d$) identically vanish due to the exact cancellation of anti-commuting Majorana pairs under independent summations. The resulting effective four-body coupling constant $\mathcal{J}^{(4)}_{\text{eff}}$ for distinct indices is explicitly expressed as:
\begin{equation}
\mathcal{J}^{(4)}_{\text{eff}} = 2 \left( 
\mathcal{J}_{a b}^{(2)}\mathcal{J}_{c d}^{(2)} + 
\mathcal{J}_{a d}^{(2)}\mathcal{J}_{b c}^{(2)} - 
\mathcal{J}_{a c}^{(2)}\mathcal{J}_{b d}^{(2)} \right).
\label{eq:J4_eff}
\end{equation}
To reproduce the disorder distribution, we choose the two-body couplings as independent Gaussian random variables with zero mean and variance $J^2 / N^3$. While this construction successfully reproduces the exact zero mean and variance of the original four-body SYK model, the resulting formulation should be viewed as a statistically equivalent representation rather than one strictly identical Hamiltonian \cite{Instability_of_SYK}. 

Following the standard DQMC method \cite{gubernatis2016}, we introduce scalar auxiliary fields $\phi_n$ at each imaginary-time slice, and define an effective action as:
\begin{equation}
S_{\text{eff}} = 
\sum_{n=1}^{N_\tau} \phi_{n}^2 - 
\ln \left( \text{sgn} \sqrt{ \text{Det}\left\{ \mathbf{I} + \prod_{n=1}^{N_\tau} \mathbf{B}_n \right\} } \right),
\label{eq:S_eff}
\end{equation}
where $\mathbf{I}$ is the identity matrix, and the propagator is defined as $\mathbf{B}_n = \exp(\phi_n \sqrt{\Delta \tau} \mathbf{K}_n)$, in which the skew-symmetric matrix is constructed with elements $\mathbf{K}_{n,ab} = 4\mathcal{J}_{n,ab}^{(2)}$. The discrete factor $\text{sgn} = \pm 1$ represents the residual sign ambiguity in the square root of Majorana fermion determinant, while this sign can be rigorously identified utilizing the Pfaffian formula introduced in Refs.~\cite{pfQMC, note_on_paired_fermions}:
\begin{equation}
\begin{aligned}
\text{Tr}\,&\prod_{n=1}^{N_\tau}
e^{\frac{1}{4}\sum_{a b}\,\chi_a \mathbf{K}_{n,a b}\chi_b}
=2^{N/2}\prod_{n=1}^{N_\tau}\rho\left( \mathbf{K}_{n} \right)\\
&\times \text{Pf}\left\{
\begin{pmatrix}
\tanh\frac{\mathbf{K}_{1}}{2}
& -\mathbf{I} & \mathbf{I} & \cdots\\
\mathbf{I} & \tanh\frac{\mathbf{K}_{2}}{2}
& \ddots & \vdots \\
-\mathbf{I} & \ddots & \ddots & -\mathbf{I} \\
\vdots & \cdots & \mathbf{I} & \tanh\frac{\mathbf{K}_{N_\tau}}{2}
\end{pmatrix}
\right\},
\end{aligned}
\label{eq:pfaffian_1}
\end{equation}
\begin{equation}
\rho\left( \mathbf{K}_{n} \right)
=\text{Pf}\left\{
\begin{pmatrix}
\sqrt{2}\,\sinh\frac{\mathbf{K}_{n}}{4} & -\mathbf{I}\\
\mathbf{I} & \sqrt{2}\,\sinh\frac{\mathbf{K}_{n}}{4}\\
\end{pmatrix}
\right\}.
\label{eq:pfaffian_2}
\end{equation}
For multiple thimbles, the relative signs are generally not identical \cite{lf_and_stochastic_q, lf_mc_many_body}, thus the exact Pfaffian formula remains indispensable for global sampling. However, the continuity of the Lefschetz thimble ensures that the sign remains invariant within any individual thimble. This local invariance allows us to safely adopt Eq.~\eqref{eq:S_eff} to a single thimble and accelerate the integration algorithm.

To validate the numerical results obtained from the Lefschetz-thimble method, we first examine the high-temperature limit $\beta \to 0$. In this regime, the propagator $\mathbf{B}_n$ approaches the identity matrix and the skew-symmetric matrix $\mathbf{K}_n$ approaches the zero matrix, justifying a positive $\text{sgn} = +1$. We expand the auxiliary field $\phi_n$ to the second order (corresponding to the first order of $\Delta\tau$) and perform the high-temperature expansion of the partition function:
\begin{equation}
\left<\Omega\right> \sim 2^{N/2} \left[ 1 - \beta J^2\frac{N-1}{N^2} + \mathcal{O} \left( \frac{\beta^2 J^4}{N^2} \right) \right],
\label{eq:partition_function_HT}
\end{equation}
where $\left< \dots \right>$ denotes the ensemble average over the Gaussian random couplings, and $\mathcal{O}$ denotes the magnitude of truncation error. Conversely, in the infrared limit characterized by low temperatures and strong couplings, we follow the ideas in Ref.~\cite{remarks_on_SYK}. The free Majorana fermion propagator becomes negligible in the Schwinger-Dyson equations, thus the Green function $\text{G}(\tau)$ emerges as an invariant under reparametrizations. This property‌ constrains the infrared-limit Green function to this ansatz:
\begin{equation}
\text{G}_{\text{IR}}(\tau) = \frac{1}{2J}\left[ \frac{1}{\beta \sin\frac{\pi \tau}{\beta}} \right]^{\frac{1}{2}}, \tau>0.
\label{eq:green_function_IR}
\end{equation}

\section{\label{sec:results}Results of Lefschetz Thimble}
Our simulation program consists of four main stages. First, we analytically continue the real auxiliary fields $\phi_n$ to complex variables $z_n$, and deploy the complex Langevin equation to search for the dominant saddle points within the complex space. Second, we utilize these saddle points as initial states, and generate thimble paths with the standard gradient flow equations. Third, we perform high-dimensional Monte Carlo integration over the generated thimbles. Due to the thimble-length constraints and the exponential decay behavior of localized weight $e^{-\text{Re}(S_{\text{eff}})}$ on the thimble, we implement the importance sampling and truncate at a boundary where the weight drops to $e^{-18}$ (a $6\sigma$ threshold). Finally, we update the random couplings $\mathcal{J}_{ab}^{(2)}$ and average over the integrated observables in the outer loop.

\subsection{\label{sec:partition_function_HT} Partition Function \\in the High-Temperature Limit}
We evaluate the scaled partition function $2^{-N/2}\langle \Omega \rangle$ in the high-temperature regime, where lower disorder fluctuations with weaker interaction strength $J$ and larger system size $N$ ensure a better numerical convergence. In the first stage, the complex Langevin search reveals that the path integral is dominated by a major saddle point located at the origin. The resulting Lefschetz thimble collapses onto the real axis, eliminating the sign ambiguity with $\text{sgn} = +1$. In the third stage, the propagator matrices $\mathbf{B}_n$ approximately degenerate into the identity matrix, thus the path integral $\Omega \propto \int d^{N_{\tau}}\phi_ne^{-S_{\text{eff}}}$ approximately degenerates into a non-interacting Gaussian integration. Consequently, the high-dimensional Monte Carlo integration achieves numerical stability, requiring only $10^3$ samplings to yield results securely within the controlled statistical error.
\begin{figure}[htbp]
\centering
\includegraphics[width=1\linewidth]{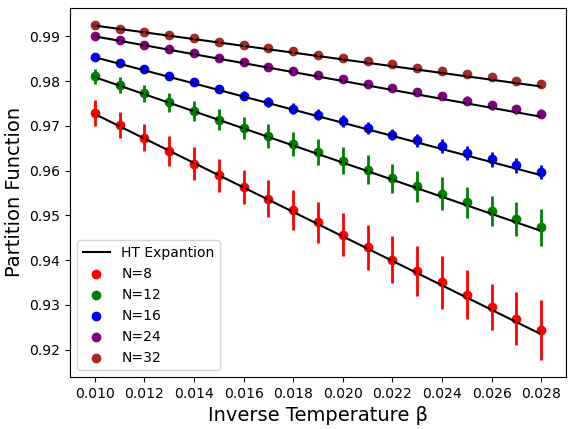}
\caption{The scaled partition function versus the inverse temperature $\beta$ for different system sizes $N$, fixed at interaction strength $J=20$ and time slice $N_\tau=16$. Vertical error bars represent the standard deviation, which become negligibly small for $N=24$ and $32$. All numerical data points are consistent with the expansion baseline within statistical errors. For complementary results with varying interaction strengths $J$, see Fig.~\ref{app_fig:ht_diff_J} in Appendix~\ref{app:partition_function_diff_J}.}
\label{fig:ht_diff_N}
\end{figure}

\subsection{Green Function in the Infrared Limit}
\begin{figure}[htbp]
\centering
\includegraphics[width=1\linewidth]{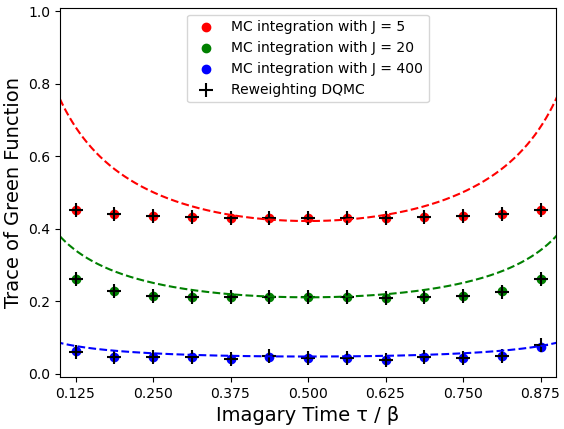}
\caption{Comparison of the imaginary-time Green function $\text{G}(\tau)$ evaluated across three frameworks fixed at $\beta=1$, $N=16$, and $N_\tau=16$. Colored dashed lines represent the infrared-limit solutions, solid markers denote the continuous Monte Carlo integrations, and black crosses indicate the discrete DQMC after reweighting ($10^4$ sampling sweeps). Both numerical schemes accurately enforce the algebraic constraint $\text{G}(0)=\text{G}(\beta)=0.5$ and display excellent agreement.}
\label{fig:mc_Reweight}
\end{figure}
As the inverse temperature $\beta$ and coupling strength $J$ increase, the complex Langevin simulation uncovers multiple relevant saddle points branching into the complex space, triggering a sign problem with complex phases. To ensure statistical precision against these intricate path configurations, we scale up our Monte Carlo integration to $10^4$ sweeps per thimble. As illustrated in Fig.~\ref{fig:mc_Reweight}, the Green function achieves reasonable agreement with the infrared-limit prediction around the mid-point $\tau = \beta/2$, which stems from the fundamental grid and boundary constraints. At $\tau = 0$, the equal-time propagator is rigidly fixed at $\text{G}(0) = 1/2$, which exactly corresponds to the free-fermion behavior rather than a divergent infrared limit. These manifest discrepancies lead to the conclusion that our current parameter configurations do not strictly satisfy the requirements of ideal infrared limit.

To provide a more conclusive verification within our finite parameter regime, we implement a discrete determinant quantum Monte Carlo (DQMC) simulation with a sign-reweighting scheme. To circumvent the sign problem, we define the absolute value of $e^{-S_{\text{eff}}}$ as a positive-definite real weight and update the corresponding sign. This redefined complex sign accounts for both the original sign ambiguity and the imaginary phases arising from the square root of the determinant. As shown in Fig.~\ref{fig:mc_Reweight}, the DQMC results are quantitatively consistent with the high-dimensional Monte Carlo integration, with their deviations falling within the statistical error bars. These two approaches match exactly at the boundaries, yielding $\text{G}(0)=\text{G}(\beta)=1/2$. Remarkably, we directly construct the Markov chain without fast update algorithms, thus this consistency stems from the fact that these two approaches differ only by the continuous or discrete form of the auxiliary fields. Furthermore, as demonstrated in Appendix~\ref{app:sign_problem}, the average redefined sign remains close to 1, indicating that the sign problem is remarkably mild within our finite parameter regime.

\subsection{Contributions of Distinct Saddle Points}
\begin{figure}[htbp]
\centering
\includegraphics[width=1\linewidth]{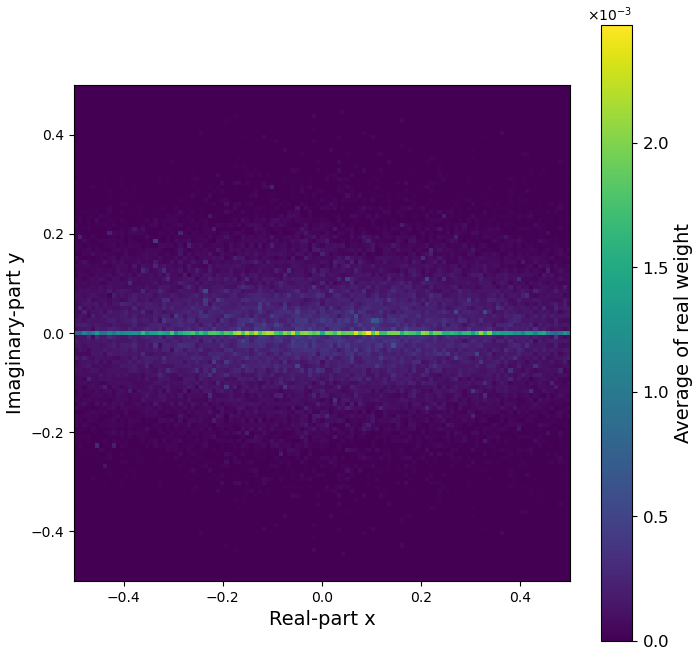}
\caption{Distribution of the real part of the average weight on the $n_\tau=1$ complex plane with $\beta=20$, $J=20$, $N=16$, and $N_\tau=16$. The color scale indicates the magnitude of the average weight, where bright yellow represents relatively large values and dark blue represents smaller values. Near the origin, the real part of the average weight is approximately $2.3\%$, which is significantly larger than the other saddle points shown in the plot. Thus the origin is omitted for clarity.}
\label{fig:saddle_weight_real}
\end{figure}

\begin{figure}[htbp]
\centering
\includegraphics[width=1\linewidth]{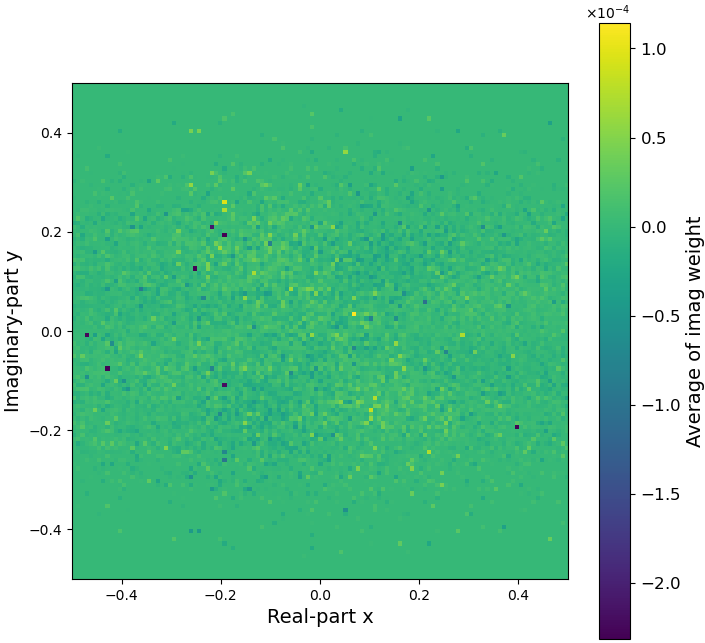}
\caption{Distribution of the imaginary part of the average weight on the $n_\tau=1$ complex plane with $\beta=20$, $J=20$, $N=16$, and $N_\tau=16$. The color scale indicates the magnitude of the average weight, where bright yellow signifies positive values, dark blue signifies negative values, and the indistinct green regions indicate values close to zero.}
\label{fig:saddle_weight_imag}
\end{figure}

The Lefschetz-thimble method suffers from residual sign problems associated with the saddle points. To address it, we employ Langevin simulations to search for a subset of contributing saddle points. The validity of our Langevin search is verified through its consistency with both the high-temperature and infrared limits. Fig.~\ref{fig:saddle_weight_real} and Fig.~\ref{fig:saddle_weight_imag} statistically visualize the complex weights between distinct saddle points across $10^3$ random matrices. The region near the origin dominates the distribution, yielding the largest real weight of about $2.3\%$. The second significant saddle points align with the real axis, with the real weights on the order of $10^{-3}$, while the other saddle points extending into the complex plane suppress rapidly to $10^{-4}$. Concurrently, the imaginary weights vanish along the real axis and exhibit only featureless stochastic noise elsewhere. These weight distributions underscore that the path integral contributions for the four-body SYK model are heavily concentrated near the real axis. Ultimately, these results confirm that our Langevin search successfully captures the relevant dominant saddle points governing the four-body SYK model in the large-$N$ limit.

\section{\label{sec:conclusions}Conclusions}
In this paper, we have implemented the Lefschetz-thimble framework in a statistically equivalent four-body SYK model, validating our numerical results against the high-temperature expansions, the infrared-limit solutions, and the discrete DQMC method. Analysis of complex weight distributions confirms that the path integral contributions are heavily concentrated near the real axis, demonstrating that the relevant saddle points govern the four-body SYK model in the large-$N$ limit.

In addition to these findings, we outline several prospects for future research. First, machine-learning algorithms or thimble-based Langevin dynamics can be integrated to overcome the computational restriction at larger system size, extending the method to stronger couplings and lower temperatures. Second, tracking the complex saddle-point trajectories vesus $\beta$, $J$, and $N$ may unveil their direct connections to quantum phase transition and chaotic behavior. Third, finite-size corrections and non-conformal crossover behaviors will help bridge the remaining gaps between numerical results and analytical theories. Finally, extending this framework to more complex sign-problematic systems, such as the two-dimensional Hubbard model, the $t$-$J$ model, and finite-density gauge theories, presents an exciting frontier. This work lays a viable framework for addressing these more complex strongly correlated systems.

\begin{acknowledgments}
Numerical computations in this work were supported by several open-source libraries. We gratefully acknowledge the use of the GNU Scientific Library (GSL) \cite{gsl_ode}, specifically utilizing its eighth-order adaptive-step Runge-Kutta algorithm for constructing the thimble and anti-thimble paths. We also acknowledge the use of the pfapack library \cite{pfapack} for the efficient numerical evaluation of the Pfaffian values.
\end{acknowledgments}

\appendix
\begin{figure}[htbp]
\centering
\includegraphics[width=1\linewidth]{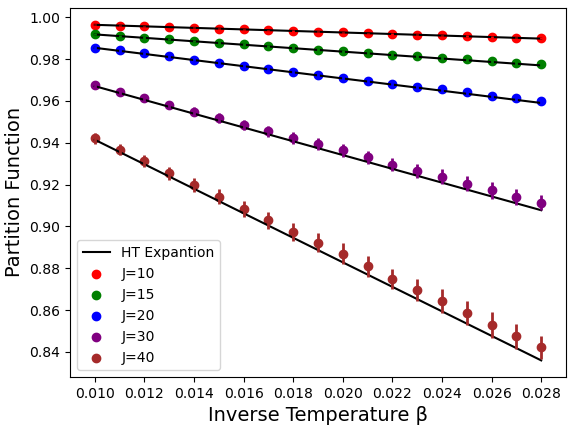}
\caption{The scaled partition function versus the inverse temperature $\beta$ for different interaction strengths $J$, fixed at system size $N=16$ and time slice $N_\tau=16$. For moderate couplings ($J \leq 20$), the error bars are tightly bounded. As both the interaction strength $J$ and the inverse temperature $\beta$ increase, the numerical results gradually deviate from the linear high-temperature expansions.}  
\label{app_fig:ht_diff_J}
\end{figure}

\section{\label{app:partition_function_diff_J} Partition Function for Different Interaction Strength}
The systematic deviation of the Monte Carlo integration results to  the high-temperature expansions in Fig.~\ref{app_fig:ht_diff_J} stems from the second-order truncation. Specifically, the Gaussian integrals of the odd-order expansions for the square root of the determinant identically vanish, and the even-order non-zero expansions possess positive coefficients. Consequently, the linear expansion inherently yields a lower bound compared to the exact, which accounts for the observed systematic deviations.

\section{\label{app:sign_problem} The Sign Problem in DQMC}
\begin{figure}[htbp]
\centering
\includegraphics[width=1\linewidth]{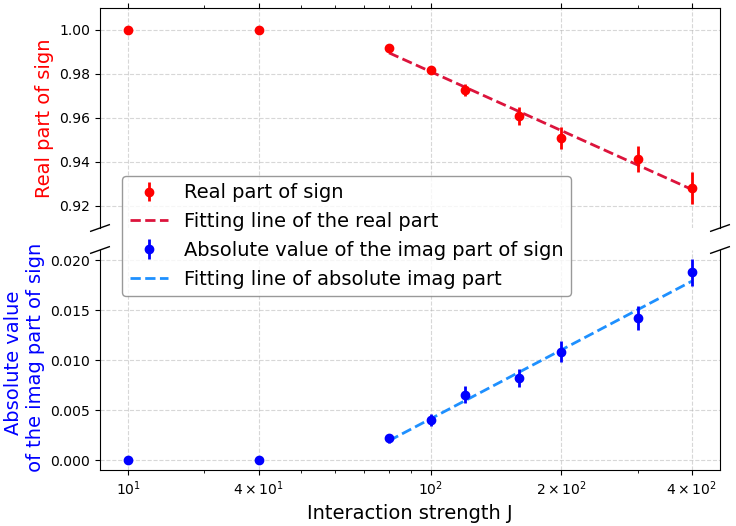}
\caption{Average of the redefined sign versus the interaction strength $J$ with $10^4$ DQMC sampling steps, fixed at $\beta=1$, $N=16$, and $N_\tau=16$. The red and blue solid markers denote the DQMC results, while the crimson and dodgerblue dashed lines represent the linear fits on a semi-logarithmic scale. No severe sign problem is observed for $J \le 40$.}
\label{app:DQMC_sign}
\end{figure}
Although DQMC method is generally susceptible to the sign problem, our saddle-point distributions suggest that its severity increases at larger $\beta$, stronger $J$, and smaller $N$. As illustrated in Fig.~\ref{app:DQMC_sign}, within our finite parameter regime, the real part of the average sign remains close to unity while the imaginary part is negligible. It confirms that the sign problem is remarkably mild, preventing substantial inflation of statistical errors in the Green functions. Conversely, the exponential behavior captured by the linear fits on the semi-logarithmic scale portends that the sign problem will eventually exacerbate drastically in the strong-coupling limit.

\bibliography{main}

\begin{thebibliography}{47}%
\makeatletter
\providecommand \@ifxundefined [1]{%
 \@ifx{#1\undefined}
}%
\providecommand \@ifnum [1]{%
 \ifnum #1\expandafter \@firstoftwo
 \else \expandafter \@secondoftwo
 \fi
}%
\providecommand \@ifx [1]{%
 \ifx #1\expandafter \@firstoftwo
 \else \expandafter \@secondoftwo
 \fi
}%
\providecommand \natexlab [1]{#1}%
\providecommand \enquote  [1]{``#1''}%
\providecommand \bibnamefont  [1]{#1}%
\providecommand \bibfnamefont [1]{#1}%
\providecommand \citenamefont [1]{#1}%
\providecommand \href@noop [0]{\@secondoftwo}%
\providecommand \href [0]{\begingroup \@sanitize@url \@href}%
\providecommand \@href[1]{\@@startlink{#1}\@@href}%
\providecommand \@@href[1]{\endgroup#1\@@endlink}%
\providecommand \@sanitize@url [0]{\catcode `\\12\catcode `\$12\catcode `\&12\catcode `\#12\catcode `\^12\catcode `\_12\catcode `\%12\relax}%
\providecommand \@@startlink[1]{}%
\providecommand \@@endlink[0]{}%
\providecommand \url  [0]{\begingroup\@sanitize@url \@url }%
\providecommand \@url [1]{\endgroup\@href {#1}{\urlprefix }}%
\providecommand \urlprefix  [0]{URL }%
\providecommand \Eprint [0]{\href }%
\providecommand \doibase [0]{https://doi.org/}%
\providecommand \selectlanguage [0]{\@gobble}%
\providecommand \bibinfo  [0]{\@secondoftwo}%
\providecommand \bibfield  [0]{\@secondoftwo}%
\providecommand \translation [1]{[#1]}%
\providecommand \BibitemOpen [0]{}%
\providecommand \bibitemStop [0]{}%
\providecommand \bibitemNoStop [0]{.\EOS\space}%
\providecommand \EOS [0]{\spacefactor3000\relax}%
\providecommand \BibitemShut  [1]{\csname bibitem#1\endcsname}%
\let\auto@bib@innerbib\@empty
\bibitem [{\citenamefont {Feynman}\ \emph {et~al.}(2010)\citenamefont {Feynman}, \citenamefont {Hibbs},\ and\ \citenamefont {Styer}}]{feynman_path}%
  \BibitemOpen
  \bibfield  {author} {\bibinfo {author} {\bibfnamefont {R.~P.}\ \bibnamefont {Feynman}}, \bibinfo {author} {\bibfnamefont {A.~R.}\ \bibnamefont {Hibbs}},\ and\ \bibinfo {author} {\bibfnamefont {D.~F.}\ \bibnamefont {Styer}},\ }\href@noop {} {\emph {\bibinfo {title} {Quantum mechanics and path integrals}}}\ (\bibinfo  {publisher} {Courier Corporation},\ \bibinfo {year} {2010})\BibitemShut {NoStop}%
\bibitem [{\citenamefont {Weinberg}(1995)}]{weinbergQFT}%
  \BibitemOpen
  \bibfield  {author} {\bibinfo {author} {\bibfnamefont {S.}~\bibnamefont {Weinberg}},\ }\href@noop {} {\emph {\bibinfo {title} {The quantum theory of fields}}},\ Vol.~\bibinfo {volume} {2}\ (\bibinfo  {publisher} {Cambridge university press},\ \bibinfo {year} {1995})\BibitemShut {NoStop}%
\bibitem [{\citenamefont {Peskin}(2018)}]{peskin2018}%
  \BibitemOpen
  \bibfield  {author} {\bibinfo {author} {\bibfnamefont {M.~E.}\ \bibnamefont {Peskin}},\ }\href@noop {} {\emph {\bibinfo {title} {An Introduction to quantum field theory}}}\ (\bibinfo  {publisher} {CRC press},\ \bibinfo {year} {2018})\BibitemShut {NoStop}%
\bibitem [{\citenamefont {Kardar}(2007)}]{kardar2007}%
  \BibitemOpen
  \bibfield  {author} {\bibinfo {author} {\bibfnamefont {M.}~\bibnamefont {Kardar}},\ }\href@noop {} {\emph {\bibinfo {title} {Statistical physics of fields}}}\ (\bibinfo  {publisher} {Cambridge University Press},\ \bibinfo {year} {2007})\BibitemShut {NoStop}%
\bibitem [{\citenamefont {Altland}\ and\ \citenamefont {Simons}(2010)}]{altland2010}%
  \BibitemOpen
  \bibfield  {author} {\bibinfo {author} {\bibfnamefont {A.}~\bibnamefont {Altland}}\ and\ \bibinfo {author} {\bibfnamefont {B.~D.}\ \bibnamefont {Simons}},\ }\href@noop {} {\emph {\bibinfo {title} {Condensed matter field theory}}}\ (\bibinfo  {publisher} {Cambridge university press},\ \bibinfo {year} {2010})\BibitemShut {NoStop}%
\bibitem [{\citenamefont {Gattringer}\ and\ \citenamefont {Lang}(2009)}]{gattringer2009}%
  \BibitemOpen
  \bibfield  {author} {\bibinfo {author} {\bibfnamefont {C.}~\bibnamefont {Gattringer}}\ and\ \bibinfo {author} {\bibfnamefont {C.}~\bibnamefont {Lang}},\ }\href@noop {} {\emph {\bibinfo {title} {Quantum chromodynamics on the lattice: an introductory presentation}}},\ Vol.\ \bibinfo {volume} {788}\ (\bibinfo  {publisher} {Springer Science \& Business Media},\ \bibinfo {year} {2009})\BibitemShut {NoStop}%
\bibitem [{\citenamefont {Blankenbecler}\ \emph {et~al.}(1981)\citenamefont {Blankenbecler}, \citenamefont {Scalapino},\ and\ \citenamefont {Sugar}}]{mc_coupled_bf}%
  \BibitemOpen
  \bibfield  {author} {\bibinfo {author} {\bibfnamefont {R.}~\bibnamefont {Blankenbecler}}, \bibinfo {author} {\bibfnamefont {D.~J.}\ \bibnamefont {Scalapino}},\ and\ \bibinfo {author} {\bibfnamefont {R.~L.}\ \bibnamefont {Sugar}},\ }\bibfield  {title} {\bibinfo {title} {Monte carlo calculations of coupled boson-fermion systems. i},\ }\href {https://doi.org/10.1103/PhysRevD.24.2278} {\bibfield  {journal} {\bibinfo  {journal} {Phys. Rev. D}\ }\textbf {\bibinfo {volume} {24}},\ \bibinfo {pages} {2278} (\bibinfo {year} {1981})}\BibitemShut {NoStop}%
\bibitem [{\citenamefont {Hirsch}(1983)}]{discrete_hs}%
  \BibitemOpen
  \bibfield  {author} {\bibinfo {author} {\bibfnamefont {J.~E.}\ \bibnamefont {Hirsch}},\ }\bibfield  {title} {\bibinfo {title} {Discrete hubbard-stratonovich transformation for fermion lattice models},\ }\href {https://doi.org/10.1103/PhysRevB.28.4059} {\bibfield  {journal} {\bibinfo  {journal} {Phys. Rev. B}\ }\textbf {\bibinfo {volume} {28}},\ \bibinfo {pages} {4059} (\bibinfo {year} {1983})}\BibitemShut {NoStop}%
\bibitem [{\citenamefont {Wei}\ \emph {et~al.}(2016)\citenamefont {Wei}, \citenamefont {Wu}, \citenamefont {Li}, \citenamefont {Zhang},\ and\ \citenamefont {Xiang}}]{Majorana_fermi_sign}%
  \BibitemOpen
  \bibfield  {author} {\bibinfo {author} {\bibfnamefont {Z.~C.}\ \bibnamefont {Wei}}, \bibinfo {author} {\bibfnamefont {C.}~\bibnamefont {Wu}}, \bibinfo {author} {\bibfnamefont {Y.}~\bibnamefont {Li}}, \bibinfo {author} {\bibfnamefont {S.}~\bibnamefont {Zhang}},\ and\ \bibinfo {author} {\bibfnamefont {T.}~\bibnamefont {Xiang}},\ }\bibfield  {title} {\bibinfo {title} {Majorana positivity and the fermion sign problem of quantum monte carlo simulations},\ }\href {https://doi.org/10.1103/PhysRevLett.116.250601} {\bibfield  {journal} {\bibinfo  {journal} {Phys. Rev. Lett.}\ }\textbf {\bibinfo {volume} {116}},\ \bibinfo {pages} {250601} (\bibinfo {year} {2016})}\BibitemShut {NoStop}%
\bibitem [{\citenamefont {Cristoforetti}\ \emph {et~al.}(2012{\natexlab{a}})\citenamefont {Cristoforetti}, \citenamefont {Renzo},\ and\ \citenamefont {Scorzato}}]{New_sign_QCD}%
  \BibitemOpen
  \bibfield  {author} {\bibinfo {author} {\bibfnamefont {A.~C.~M.}\ \bibnamefont {Cristoforetti}}, \bibinfo {author} {\bibfnamefont {F.~D.}\ \bibnamefont {Renzo}},\ and\ \bibinfo {author} {\bibfnamefont {L.}~\bibnamefont {Scorzato}},\ }\bibfield  {title} {\bibinfo {title} {New approach to the sign problem in quantum field theories: High density qcd on a lefschetz thimble},\ }\href {https://api.semanticscholar.org/CorpusID:118720098} {\bibfield  {journal} {\bibinfo  {journal} {Physical Review D}\ }\textbf {\bibinfo {volume} {86}},\ \bibinfo {pages} {074506} (\bibinfo {year} {2012}{\natexlab{a}})}\BibitemShut {NoStop}%
\bibitem [{\citenamefont {Corboz}\ \emph {et~al.}(2014)\citenamefont {Corboz}, \citenamefont {Rice},\ and\ \citenamefont {Troyer}}]{t_J_tensor}%
  \BibitemOpen
  \bibfield  {author} {\bibinfo {author} {\bibfnamefont {P.}~\bibnamefont {Corboz}}, \bibinfo {author} {\bibfnamefont {T.~M.}\ \bibnamefont {Rice}},\ and\ \bibinfo {author} {\bibfnamefont {M.}~\bibnamefont {Troyer}},\ }\bibfield  {title} {\bibinfo {title} {Competing states in the $t$-$j$ model: Uniform $d$-wave state versus stripe state},\ }\href {https://doi.org/10.1103/PhysRevLett.113.046402} {\bibfield  {journal} {\bibinfo  {journal} {Phys. Rev. Lett.}\ }\textbf {\bibinfo {volume} {113}},\ \bibinfo {pages} {046402} (\bibinfo {year} {2014})}\BibitemShut {NoStop}%
\bibitem [{\citenamefont {Wan}\ \emph {et~al.}(2022)\citenamefont {Wan}, \citenamefont {Zhang},\ and\ \citenamefont {Yao}}]{Fermion_S_auto_dif}%
  \BibitemOpen
  \bibfield  {author} {\bibinfo {author} {\bibfnamefont {Z.-Q.}\ \bibnamefont {Wan}}, \bibinfo {author} {\bibfnamefont {S.-X.}\ \bibnamefont {Zhang}},\ and\ \bibinfo {author} {\bibfnamefont {H.}~\bibnamefont {Yao}},\ }\bibfield  {title} {\bibinfo {title} {Mitigating the fermion sign problem by automatic differentiation},\ }\href {https://doi.org/10.1103/PhysRevB.106.L241109} {\bibfield  {journal} {\bibinfo  {journal} {Phys. Rev. B}\ }\textbf {\bibinfo {volume} {106}},\ \bibinfo {pages} {L241109} (\bibinfo {year} {2022})}\BibitemShut {NoStop}%
\bibitem [{\citenamefont {Li}\ \emph {et~al.}(2015)\citenamefont {Li}, \citenamefont {Jiang},\ and\ \citenamefont {Yao}}]{solve_fermion_sign}%
  \BibitemOpen
  \bibfield  {author} {\bibinfo {author} {\bibfnamefont {Z.-X.}\ \bibnamefont {Li}}, \bibinfo {author} {\bibfnamefont {Y.-F.}\ \bibnamefont {Jiang}},\ and\ \bibinfo {author} {\bibfnamefont {H.}~\bibnamefont {Yao}},\ }\bibfield  {title} {\bibinfo {title} {Solving the fermion sign problem in quantum monte carlo simulations by majorana representation},\ }\href {https://doi.org/10.1103/PhysRevB.91.241117} {\bibfield  {journal} {\bibinfo  {journal} {Phys. Rev. B}\ }\textbf {\bibinfo {volume} {91}},\ \bibinfo {pages} {241117} (\bibinfo {year} {2015})}\BibitemShut {NoStop}%
\bibitem [{\citenamefont {White}\ \emph {et~al.}(1989)\citenamefont {White}, \citenamefont {Scalapino}, \citenamefont {Sugar}, \citenamefont {Loh}, \citenamefont {Gubernatis},\ and\ \citenamefont {Scalettar}}]{study_2d_hubbard}%
  \BibitemOpen
  \bibfield  {author} {\bibinfo {author} {\bibfnamefont {S.~R.}\ \bibnamefont {White}}, \bibinfo {author} {\bibfnamefont {D.~J.}\ \bibnamefont {Scalapino}}, \bibinfo {author} {\bibfnamefont {R.~L.}\ \bibnamefont {Sugar}}, \bibinfo {author} {\bibfnamefont {E.~Y.}\ \bibnamefont {Loh}}, \bibinfo {author} {\bibfnamefont {J.~E.}\ \bibnamefont {Gubernatis}},\ and\ \bibinfo {author} {\bibfnamefont {R.~T.}\ \bibnamefont {Scalettar}},\ }\bibfield  {title} {\bibinfo {title} {Numerical study of the two-dimensional hubbard model},\ }\href {https://doi.org/10.1103/PhysRevB.40.506} {\bibfield  {journal} {\bibinfo  {journal} {Phys. Rev. B}\ }\textbf {\bibinfo {volume} {40}},\ \bibinfo {pages} {506} (\bibinfo {year} {1989})}\BibitemShut {NoStop}%
\bibitem [{\citenamefont {Gavai}(1985)}]{chemical_lattice}%
  \BibitemOpen
  \bibfield  {author} {\bibinfo {author} {\bibfnamefont {R.~V.}\ \bibnamefont {Gavai}},\ }\bibfield  {title} {\bibinfo {title} {Chemical potential on the lattice},\ }\href {https://doi.org/10.1103/PhysRevD.32.519} {\bibfield  {journal} {\bibinfo  {journal} {Phys. Rev. D}\ }\textbf {\bibinfo {volume} {32}},\ \bibinfo {pages} {519} (\bibinfo {year} {1985})}\BibitemShut {NoStop}%
\bibitem [{\citenamefont {Li}\ \emph {et~al.}(2011)\citenamefont {Li}, \citenamefont {Alexandru},\ and\ \citenamefont {Liu}}]{C_point_lattice}%
  \BibitemOpen
  \bibfield  {author} {\bibinfo {author} {\bibfnamefont {A.}~\bibnamefont {Li}}, \bibinfo {author} {\bibfnamefont {A.}~\bibnamefont {Alexandru}},\ and\ \bibinfo {author} {\bibfnamefont {K.-F.}\ \bibnamefont {Liu}},\ }\bibfield  {title} {\bibinfo {title} {Critical point of ${N}_{f}\mathbf{=}3$ qcd from lattice simulations in the canonical ensemble},\ }\href {https://doi.org/10.1103/PhysRevD.84.071503} {\bibfield  {journal} {\bibinfo  {journal} {Phys. Rev. D}\ }\textbf {\bibinfo {volume} {84}},\ \bibinfo {pages} {071503} (\bibinfo {year} {2011})}\BibitemShut {NoStop}%
\bibitem [{\citenamefont {Grossman}\ and\ \citenamefont {Berg}(2023)}]{fermi_liquid_sign_free}%
  \BibitemOpen
  \bibfield  {author} {\bibinfo {author} {\bibfnamefont {O.}~\bibnamefont {Grossman}}\ and\ \bibinfo {author} {\bibfnamefont {E.}~\bibnamefont {Berg}},\ }\bibfield  {title} {\bibinfo {title} {Robust fermi-liquid instabilities in sign problem-free models},\ }\href {https://doi.org/10.1103/PhysRevLett.131.056501} {\bibfield  {journal} {\bibinfo  {journal} {Phys. Rev. Lett.}\ }\textbf {\bibinfo {volume} {131}},\ \bibinfo {pages} {056501} (\bibinfo {year} {2023})}\BibitemShut {NoStop}%
\bibitem [{\citenamefont {Witten}(2010{\natexlab{a}})}]{Witten2010}%
  \BibitemOpen
  \bibfield  {author} {\bibinfo {author} {\bibfnamefont {E.}~\bibnamefont {Witten}},\ }\bibfield  {title} {\bibinfo {title} {A new look at the path integral of quantum mechanics},\ }\href {https://doi.org/10.4310/SDG.2010.v15.n1.a11} {\bibfield  {journal} {\bibinfo  {journal} {Surveys in Differential Geometry}\ }\textbf {\bibinfo {volume} {15}} (\bibinfo {year} {2010}{\natexlab{a}})}\BibitemShut {NoStop}%
\bibitem [{\citenamefont {Kanazawa}\ and\ \citenamefont {Tanizaki}(2014)}]{structure_lf_simple_fermion}%
  \BibitemOpen
  \bibfield  {author} {\bibinfo {author} {\bibfnamefont {T.}~\bibnamefont {Kanazawa}}\ and\ \bibinfo {author} {\bibfnamefont {Y.}~\bibnamefont {Tanizaki}},\ }\bibfield  {title} {\bibinfo {title} {Structure of lefschetz thimbles in simple fermionic systems},\ }\href {https://doi.org/10.1007/JHEP03(2015)044} {\bibfield  {journal} {\bibinfo  {journal} {Journal of High Energy Physics}\ }\textbf {\bibinfo {volume} {2015}} (\bibinfo {year} {2014})}\BibitemShut {NoStop}%
\bibitem [{\citenamefont {Fujii}\ \emph {et~al.}(2015)\citenamefont {Fujii}, \citenamefont {Kamata},\ and\ \citenamefont {Kikukawa}}]{lf_one_lattice}%
  \BibitemOpen
  \bibfield  {author} {\bibinfo {author} {\bibfnamefont {H.}~\bibnamefont {Fujii}}, \bibinfo {author} {\bibfnamefont {S.}~\bibnamefont {Kamata}},\ and\ \bibinfo {author} {\bibfnamefont {Y.}~\bibnamefont {Kikukawa}},\ }\bibfield  {title} {\bibinfo {title} {Lefschetz thimble structure in one-dimensional lattice thirring model at finite density},\ }\href {https://doi.org/10.1007/JHEP11(2015)078} {\bibfield  {journal} {\bibinfo  {journal} {Journal of High Energy Physics}\ }\textbf {\bibinfo {volume} {2015}} (\bibinfo {year} {2015})}\BibitemShut {NoStop}%
\bibitem [{\citenamefont {Tanizaki}\ \emph {et~al.}(2016)\citenamefont {Tanizaki}, \citenamefont {Hidaka},\ and\ \citenamefont {Hayata}}]{lef_of_sign}%
  \BibitemOpen
  \bibfield  {author} {\bibinfo {author} {\bibfnamefont {Y.}~\bibnamefont {Tanizaki}}, \bibinfo {author} {\bibfnamefont {Y.}~\bibnamefont {Hidaka}},\ and\ \bibinfo {author} {\bibfnamefont {T.}~\bibnamefont {Hayata}},\ }\bibfield  {title} {\bibinfo {title} {Lefschetz-thimble analysis of the sign problem in one-site fermion model},\ }\href {https://doi.org/10.1088/1367-2630/18/3/033002} {\bibfield  {journal} {\bibinfo  {journal} {New Journal of Physics}\ }\textbf {\bibinfo {volume} {18}},\ \bibinfo {pages} {033002} (\bibinfo {year} {2016})}\BibitemShut {NoStop}%
\bibitem [{\citenamefont {Cristoforetti}\ \emph {et~al.}(2012{\natexlab{b}})\citenamefont {Cristoforetti}, \citenamefont {Di~Renzo},\ and\ \citenamefont {Scorzato}}]{new_approach_to_s}%
  \BibitemOpen
  \bibfield  {author} {\bibinfo {author} {\bibfnamefont {M.}~\bibnamefont {Cristoforetti}}, \bibinfo {author} {\bibfnamefont {F.}~\bibnamefont {Di~Renzo}},\ and\ \bibinfo {author} {\bibfnamefont {L.}~\bibnamefont {Scorzato}},\ }\bibfield  {title} {\bibinfo {title} {New approach to the sign problem in quantum field theories: High density qcd on a lefschetz thimble},\ }\href {https://doi.org/10.1103/PhysRevD.86.074506} {\bibfield  {journal} {\bibinfo  {journal} {Phys. Rev. D}\ }\textbf {\bibinfo {volume} {86}} (\bibinfo {year} {2012}{\natexlab{b}})}\BibitemShut {NoStop}%
\bibitem [{\citenamefont {Cristoforetti}\ \emph {et~al.}(2013)\citenamefont {Cristoforetti}, \citenamefont {Di~Renzo}, \citenamefont {Mukherjee},\ and\ \citenamefont {Scorzato}}]{mc_on_lef}%
  \BibitemOpen
  \bibfield  {author} {\bibinfo {author} {\bibfnamefont {M.}~\bibnamefont {Cristoforetti}}, \bibinfo {author} {\bibfnamefont {F.}~\bibnamefont {Di~Renzo}}, \bibinfo {author} {\bibfnamefont {A.}~\bibnamefont {Mukherjee}},\ and\ \bibinfo {author} {\bibfnamefont {L.}~\bibnamefont {Scorzato}},\ }\bibfield  {title} {\bibinfo {title} {Monte carlo simulations on the lefschetz thimble: Taming the sign problem},\ }\href {https://doi.org/10.1103/PhysRevD.88.051501} {\bibfield  {journal} {\bibinfo  {journal} {Phys. Rev. D}\ }\textbf {\bibinfo {volume} {88}},\ \bibinfo {pages} {051501} (\bibinfo {year} {2013})}\BibitemShut {NoStop}%
\bibitem [{\citenamefont {Alexandru}\ \emph {et~al.}(2015)\citenamefont {Alexandru}, \citenamefont {Basar}, \citenamefont {Bedaque}, \citenamefont {Ridgway},\ and\ \citenamefont {Warrington}}]{sign_and_mc_beyond_lef}%
  \BibitemOpen
  \bibfield  {author} {\bibinfo {author} {\bibfnamefont {A.}~\bibnamefont {Alexandru}}, \bibinfo {author} {\bibfnamefont {G.}~\bibnamefont {Basar}}, \bibinfo {author} {\bibfnamefont {P.}~\bibnamefont {Bedaque}}, \bibinfo {author} {\bibfnamefont {G.}~\bibnamefont {Ridgway}},\ and\ \bibinfo {author} {\bibfnamefont {N.}~\bibnamefont {Warrington}},\ }\bibfield  {title} {\bibinfo {title} {Sign problem and monte carlo calculations beyond lefschetz thimbles},\ }\href {https://doi.org/10.1007/JHEP05(2016)053} {\bibfield  {journal} {\bibinfo  {journal} {Journal of High Energy Physics}\ }\textbf {\bibinfo {volume} {2016}} (\bibinfo {year} {2015})}\BibitemShut {NoStop}%
\bibitem [{\citenamefont {Born}\ and\ \citenamefont {Wolf}(2013)}]{born2013}%
  \BibitemOpen
  \bibfield  {author} {\bibinfo {author} {\bibfnamefont {M.}~\bibnamefont {Born}}\ and\ \bibinfo {author} {\bibfnamefont {E.}~\bibnamefont {Wolf}},\ }\href@noop {} {\emph {\bibinfo {title} {Principles of optics: electromagnetic theory of propagation, interference and diffraction of light}}}\ (\bibinfo  {publisher} {Elsevier},\ \bibinfo {year} {2013})\BibitemShut {NoStop}%
\bibitem [{\citenamefont {Feldbrugge}\ \emph {et~al.}(2025)\citenamefont {Feldbrugge}, \citenamefont {Jow},\ and\ \citenamefont {Pen}}]{c_quantum_reflections}%
  \BibitemOpen
  \bibfield  {author} {\bibinfo {author} {\bibfnamefont {J.}~\bibnamefont {Feldbrugge}}, \bibinfo {author} {\bibfnamefont {D.~L.}\ \bibnamefont {Jow}},\ and\ \bibinfo {author} {\bibfnamefont {U.-L.}\ \bibnamefont {Pen}},\ }\bibfield  {title} {\bibinfo {title} {Complex classical paths in quantum reflections and tunneling},\ }\href {https://doi.org/10.1103/PhysRevD.111.085027} {\bibfield  {journal} {\bibinfo  {journal} {Phys. Rev. D}\ }\textbf {\bibinfo {volume} {111}},\ \bibinfo {pages} {085027} (\bibinfo {year} {2025})}\BibitemShut {NoStop}%
\bibitem [{\citenamefont {Feldbrugge}\ \emph {et~al.}(2017)\citenamefont {Feldbrugge}, \citenamefont {Lehners},\ and\ \citenamefont {Turok}}]{Lorentzian_q_cosmplogy}%
  \BibitemOpen
  \bibfield  {author} {\bibinfo {author} {\bibfnamefont {J.}~\bibnamefont {Feldbrugge}}, \bibinfo {author} {\bibfnamefont {J.-L.}\ \bibnamefont {Lehners}},\ and\ \bibinfo {author} {\bibfnamefont {N.}~\bibnamefont {Turok}},\ }\bibfield  {title} {\bibinfo {title} {Lorentzian quantum cosmology},\ }\href {https://doi.org/10.1103/PhysRevD.95.103508} {\bibfield  {journal} {\bibinfo  {journal} {Phys. Rev. D}\ }\textbf {\bibinfo {volume} {95}},\ \bibinfo {pages} {103508} (\bibinfo {year} {2017})}\BibitemShut {NoStop}%
\bibitem [{\citenamefont {Perez}(2012)}]{spin_foam_q_gravity}%
  \BibitemOpen
  \bibfield  {author} {\bibinfo {author} {\bibfnamefont {A.}~\bibnamefont {Perez}},\ }\bibfield  {title} {\bibinfo {title} {The spin-foam approach to quantum gravity},\ }\href {https://api.semanticscholar.org/CorpusID:18022565} {\bibfield  {journal} {\bibinfo  {journal} {Living Reviews in Relativity}\ }\textbf {\bibinfo {volume} {16}} (\bibinfo {year} {2012})}\BibitemShut {NoStop}%
\bibitem [{\citenamefont {Don{\'a}}\ \emph {et~al.}(2019)\citenamefont {Don{\'a}}, \citenamefont {Fanizza}, \citenamefont {Sarno},\ and\ \citenamefont {Speziale}}]{numerical_spin_foam}%
  \BibitemOpen
  \bibfield  {author} {\bibinfo {author} {\bibfnamefont {P.}~\bibnamefont {Don{\'a}}}, \bibinfo {author} {\bibfnamefont {M.}~\bibnamefont {Fanizza}}, \bibinfo {author} {\bibfnamefont {G.}~\bibnamefont {Sarno}},\ and\ \bibinfo {author} {\bibfnamefont {S.}~\bibnamefont {Speziale}},\ }\bibfield  {title} {\bibinfo {title} {Numerical study of the lorentzian engle-pereira-rovelli-livine spin foam amplitude},\ }\href {https://doi.org/10.1103/PhysRevD.100.106003} {\bibfield  {journal} {\bibinfo  {journal} {Phys. Rev. D}\ }\textbf {\bibinfo {volume} {100}},\ \bibinfo {pages} {106003} (\bibinfo {year} {2019})}\BibitemShut {NoStop}%
\bibitem [{\citenamefont {Han}\ \emph {et~al.}(2021)\citenamefont {Han}, \citenamefont {Huang}, \citenamefont {Liu}, \citenamefont {Qu},\ and\ \citenamefont {Wan}}]{spinform_lf_mc}%
  \BibitemOpen
  \bibfield  {author} {\bibinfo {author} {\bibfnamefont {M.}~\bibnamefont {Han}}, \bibinfo {author} {\bibfnamefont {Z.}~\bibnamefont {Huang}}, \bibinfo {author} {\bibfnamefont {H.}~\bibnamefont {Liu}}, \bibinfo {author} {\bibfnamefont {D.}~\bibnamefont {Qu}},\ and\ \bibinfo {author} {\bibfnamefont {Y.}~\bibnamefont {Wan}},\ }\bibfield  {title} {\bibinfo {title} {Spinfoam on a lefschetz thimble: Markov chain monte carlo computation of a lorentzian spinfoam propagator},\ }\href {https://doi.org/10.1103/PhysRevD.103.084026} {\bibfield  {journal} {\bibinfo  {journal} {Phys. Rev. D}\ }\textbf {\bibinfo {volume} {103}},\ \bibinfo {pages} {084026} (\bibinfo {year} {2021})}\BibitemShut {NoStop}%
\bibitem [{\citenamefont {Fujii}\ \emph {et~al.}(2013)\citenamefont {Fujii}, \citenamefont {Honda}, \citenamefont {Kato}, \citenamefont {Kikukawa}, \citenamefont {Komatsu},\ and\ \citenamefont {Sano}}]{hybrid_mc_on_lef}%
  \BibitemOpen
  \bibfield  {author} {\bibinfo {author} {\bibfnamefont {H.}~\bibnamefont {Fujii}}, \bibinfo {author} {\bibfnamefont {D.}~\bibnamefont {Honda}}, \bibinfo {author} {\bibfnamefont {M.}~\bibnamefont {Kato}}, \bibinfo {author} {\bibfnamefont {Y.}~\bibnamefont {Kikukawa}}, \bibinfo {author} {\bibfnamefont {S.}~\bibnamefont {Komatsu}},\ and\ \bibinfo {author} {\bibfnamefont {T.}~\bibnamefont {Sano}},\ }\bibfield  {title} {\bibinfo {title} {Hybrid monte carlo on lefschetz thimbles - a study of the residual sign problem},\ }\href {https://api.semanticscholar.org/CorpusID:119221705} {\bibfield  {journal} {\bibinfo  {journal} {Journal of High Energy Physics}\ }\textbf {\bibinfo {volume} {2013}},\ \bibinfo {pages} {1} (\bibinfo {year} {2013})}\BibitemShut {NoStop}%
\bibitem [{\citenamefont {Tsutsui}\ and\ \citenamefont {Doi}(2016)}]{complex_langevin_lef}%
  \BibitemOpen
  \bibfield  {author} {\bibinfo {author} {\bibfnamefont {S.}~\bibnamefont {Tsutsui}}\ and\ \bibinfo {author} {\bibfnamefont {T.~M.}\ \bibnamefont {Doi}},\ }\bibfield  {title} {\bibinfo {title} {Improvement in complex langevin dynamics from a view point of lefschetz thimbles},\ }\href {https://doi.org/10.1103/PhysRevD.94.074009} {\bibfield  {journal} {\bibinfo  {journal} {Phys. Rev. D}\ }\textbf {\bibinfo {volume} {94}},\ \bibinfo {pages} {074009} (\bibinfo {year} {2016})}\BibitemShut {NoStop}%
\bibitem [{\citenamefont {Pham}(2013)}]{pham2013}%
  \BibitemOpen
  \bibfield  {author} {\bibinfo {author} {\bibfnamefont {F.}~\bibnamefont {Pham}},\ }\href@noop {} {\emph {\bibinfo {title} {Singularit{\'e}s des systemes diff{\'e}rentiels de Gauss-Manin}}}\ (\bibinfo  {publisher} {Springer Science \& Business Media},\ \bibinfo {year} {2013})\BibitemShut {NoStop}%
\bibitem [{\citenamefont {Witten}(2010{\natexlab{b}})}]{chern_simons}%
  \BibitemOpen
  \bibfield  {author} {\bibinfo {author} {\bibfnamefont {E.}~\bibnamefont {Witten}},\ }\bibfield  {title} {\bibinfo {title} {Analytic continuation of chern-simons theory},\ }\href@noop {} {\bibfield  {journal} {\bibinfo  {journal} {AMS/IP Stud. Adv. Math.}\ }\textbf {\bibinfo {volume} {50}} (\bibinfo {year} {2010}{\natexlab{b}})}\BibitemShut {NoStop}%
\bibitem [{\citenamefont {Aarts}(2009)}]{Aarts_langevin}%
  \BibitemOpen
  \bibfield  {author} {\bibinfo {author} {\bibfnamefont {G.}~\bibnamefont {Aarts}},\ }\bibfield  {title} {\bibinfo {title} {Can stochastic quantization evade the sign problem? the relativistic bose gas at finite chemical potential},\ }\href {https://doi.org/10.1103/PhysRevLett.102.131601} {\bibfield  {journal} {\bibinfo  {journal} {Phys. Rev. Lett.}\ }\textbf {\bibinfo {volume} {102}},\ \bibinfo {pages} {131601} (\bibinfo {year} {2009})}\BibitemShut {NoStop}%
\bibitem [{\citenamefont {Sachdev}\ and\ \citenamefont {Ye}(1993)}]{spin_fluid_random_H}%
  \BibitemOpen
  \bibfield  {author} {\bibinfo {author} {\bibfnamefont {S.}~\bibnamefont {Sachdev}}\ and\ \bibinfo {author} {\bibfnamefont {J.}~\bibnamefont {Ye}},\ }\bibfield  {title} {\bibinfo {title} {Gapless spin-fluid ground state in a random quantum heisenberg magnet},\ }\href {https://doi.org/10.1103/PhysRevLett.70.3339} {\bibfield  {journal} {\bibinfo  {journal} {Phys. Rev. Lett.}\ }\textbf {\bibinfo {volume} {70}},\ \bibinfo {pages} {3339} (\bibinfo {year} {1993})}\BibitemShut {NoStop}%
\bibitem [{\citenamefont {Kitaev}({\natexlab{a}})}]{kitaev1}%
  \BibitemOpen
  \bibfield  {author} {\bibinfo {author} {\bibfnamefont {A.}~\bibnamefont {Kitaev}},\ }\bibfield  {title} {\bibinfo {title} {A simple model of quantum holography (part 1)}} ({\natexlab{a}})\BibitemShut {NoStop}%
\bibitem [{\citenamefont {Kitaev}({\natexlab{b}})}]{kitaev2}%
  \BibitemOpen
  \bibfield  {author} {\bibinfo {author} {\bibfnamefont {A.}~\bibnamefont {Kitaev}},\ }\bibfield  {title} {\bibinfo {title} {A simple model of quantum holography (part 2)}} ({\natexlab{b}})\BibitemShut {NoStop}%
\bibitem [{\citenamefont {Bi}\ \emph {et~al.}(2017)\citenamefont {Bi}, \citenamefont {Jian}, \citenamefont {You}, \citenamefont {Pawlak},\ and\ \citenamefont {Xu}}]{Instability_of_SYK}%
  \BibitemOpen
  \bibfield  {author} {\bibinfo {author} {\bibfnamefont {Z.}~\bibnamefont {Bi}}, \bibinfo {author} {\bibfnamefont {C.-M.}\ \bibnamefont {Jian}}, \bibinfo {author} {\bibfnamefont {Y.-Z.}\ \bibnamefont {You}}, \bibinfo {author} {\bibfnamefont {K.~A.}\ \bibnamefont {Pawlak}},\ and\ \bibinfo {author} {\bibfnamefont {C.}~\bibnamefont {Xu}},\ }\bibfield  {title} {\bibinfo {title} {Instability of the non-fermi-liquid state of the sachdev-ye-kitaev model},\ }\href {https://doi.org/10.1103/PhysRevB.95.205105} {\bibfield  {journal} {\bibinfo  {journal} {Phys. Rev. B}\ }\textbf {\bibinfo {volume} {95}},\ \bibinfo {pages} {205105} (\bibinfo {year} {2017})}\BibitemShut {NoStop}%
\bibitem [{\citenamefont {Gubernatis}\ \emph {et~al.}(2016)\citenamefont {Gubernatis}, \citenamefont {Kawashima},\ and\ \citenamefont {Werner}}]{gubernatis2016}%
  \BibitemOpen
  \bibfield  {author} {\bibinfo {author} {\bibfnamefont {J.}~\bibnamefont {Gubernatis}}, \bibinfo {author} {\bibfnamefont {N.}~\bibnamefont {Kawashima}},\ and\ \bibinfo {author} {\bibfnamefont {P.}~\bibnamefont {Werner}},\ }\href@noop {} {\emph {\bibinfo {title} {Quantum Monte Carlo Methods: Algorithms for Lattice Models}}}\ (\bibinfo  {publisher} {Cambridge University Press},\ \bibinfo {year} {2016})\BibitemShut {NoStop}%
\bibitem [{\citenamefont {Han}\ \emph {et~al.}(2024)\citenamefont {Han}, \citenamefont {Wan},\ and\ \citenamefont {Yao}}]{pfQMC}%
  \BibitemOpen
  \bibfield  {author} {\bibinfo {author} {\bibfnamefont {Z.-Y.}\ \bibnamefont {Han}}, \bibinfo {author} {\bibfnamefont {Z.-Q.}\ \bibnamefont {Wan}},\ and\ \bibinfo {author} {\bibfnamefont {H.}~\bibnamefont {Yao}},\ }\href@noop {} {\bibinfo {title} {{Pfaffian quantum Monte Carlo: solution to Majorana sign ambiguity and applications}}} (\bibinfo {year} {2024}),\ \Eprint {https://arxiv.org/abs/2408.10311} {{arXiv}:2408.10311 [{cond-mat.str-el}]} \BibitemShut {NoStop}%
\bibitem [{\citenamefont {Klich}(2014)}]{note_on_paired_fermions}%
  \BibitemOpen
  \bibfield  {author} {\bibinfo {author} {\bibfnamefont {I.}~\bibnamefont {Klich}},\ }\bibfield  {title} {\bibinfo {title} {A note on the full counting statistics of paired fermions},\ }\href {https://doi.org/10.1088/1742-5468/2014/11/P11006} {\bibfield  {journal} {\bibinfo  {journal} {Journal of Statistical Mechanics: Theory and Experiment}\ }\textbf {\bibinfo {volume} {2014}},\ \bibinfo {pages} {P11006} (\bibinfo {year} {2014})}\BibitemShut {NoStop}%
\bibitem [{\citenamefont {Aarts}(2013)}]{lf_and_stochastic_q}%
  \BibitemOpen
  \bibfield  {author} {\bibinfo {author} {\bibfnamefont {G.}~\bibnamefont {Aarts}},\ }\bibfield  {title} {\bibinfo {title} {Lefschetz thimbles and stochastic quantization: Complex actions in the complex plane},\ }\href {https://doi.org/10.1103/PhysRevD.88.094501} {\bibfield  {journal} {\bibinfo  {journal} {Phys. Rev. D}\ }\textbf {\bibinfo {volume} {88}},\ \bibinfo {pages} {094501} (\bibinfo {year} {2013})}\BibitemShut {NoStop}%
\bibitem [{\citenamefont {Mukherjee}\ and\ \citenamefont {Cristoforetti}(2014)}]{lf_mc_many_body}%
  \BibitemOpen
  \bibfield  {author} {\bibinfo {author} {\bibfnamefont {A.}~\bibnamefont {Mukherjee}}\ and\ \bibinfo {author} {\bibfnamefont {M.}~\bibnamefont {Cristoforetti}},\ }\bibfield  {title} {\bibinfo {title} {Lefschetz thimble monte carlo for many-body theories: A hubbard model study},\ }\href {https://doi.org/10.1103/PhysRevB.90.035134} {\bibfield  {journal} {\bibinfo  {journal} {Phys. Rev. B}\ }\textbf {\bibinfo {volume} {90}},\ \bibinfo {pages} {035134} (\bibinfo {year} {2014})}\BibitemShut {NoStop}%
\bibitem [{\citenamefont {Maldacena}\ and\ \citenamefont {Stanford}(2016)}]{remarks_on_SYK}%
  \BibitemOpen
  \bibfield  {author} {\bibinfo {author} {\bibfnamefont {J.}~\bibnamefont {Maldacena}}\ and\ \bibinfo {author} {\bibfnamefont {D.}~\bibnamefont {Stanford}},\ }\bibfield  {title} {\bibinfo {title} {Remarks on the sachdev-ye-kitaev model},\ }\href {https://doi.org/10.1103/PhysRevD.94.106002} {\bibfield  {journal} {\bibinfo  {journal} {Phys. Rev. D}\ }\textbf {\bibinfo {volume} {94}},\ \bibinfo {pages} {106002} (\bibinfo {year} {2016})}\BibitemShut {NoStop}%
\bibitem [{\citenamefont {Galassi}()}]{gsl_ode}%
  \BibitemOpen
  \bibfield  {author} {\bibinfo {author} {\bibfnamefont {M.}~\bibnamefont {Galassi}},\ }\href {https://www.gnu.org/software/gsl/doc/html/ode-initval.html#stepping-functions} {\bibinfo {title} {Gnu scientific library reference manual}}\BibitemShut {NoStop}%
\bibitem [{\citenamefont {Nijholt}()}]{pfapack}%
  \BibitemOpen
  \bibfield  {author} {\bibinfo {author} {\bibfnamefont {B.}~\bibnamefont {Nijholt}},\ }\href {https://pfapack.readthedocs.io/en/latest/index.html} {\bibinfo {title} {Pfapack: Efficient numerical computation of the pfaffian}}\BibitemShut {NoStop}%
\end{thebibliography}%

\end{document}